\documentclass[journal]{IEEEtran}
\usepackage{amsmath,amsfonts}
\usepackage{algorithmic}
\usepackage{algorithm}
\usepackage{array}
\usepackage[caption=false,font=normalsize,labelfont=sf,textfont=sf]{subfig}
\usepackage{textcomp}
\usepackage{stfloats}
\usepackage{url}
\usepackage{verbatim}
\usepackage{graphicx}
\usepackage{cite}
\usepackage{multirow}
\usepackage{gensymb}

\begin{document}

\title{Trans${^2}$-CBCT: A Dual-Transformer Framework for Sparse-View CBCT Reconstruction}

\author{Minmin Yang, Huantao Ren, Senem Velipasalar

\thanks{Minmin Yang, Huantao Ren and Senem Velipasalar are with the Department of Electrical Engineering and Computer Science at Syracuse University, Syracuse, NY (email: \{myang47, hren11, svelipas\}@syr.edu).}}



\maketitle
\begin{abstract}
Cone‑beam computed tomography (CBCT) using only a handful of X-ray projection views offers the potential for significantly reduced radiation dose and faster scans. However, such extreme angular under-sampling, typically limited to six to ten views, introduces severe streak artifacts and results in large gaps in spatial coverage. These limitations significantly hinder both the extraction of reliable 2D features and the reconstruction of a coherent, artifact-suppressed 3D volume from sparse projections. In this work, we address these challenges within a unified framework. First, we replace conventional UNet/ResNet encoders with TransUNet, a hybrid CNN–Transformer architecture originally designed for medical image segmentation. The convolution layers capture fine-grained local details, while the self‑attention layers provide long‑range contextual understanding. We adapt TransUNet to CBCT by concatenating multi‑scale feature maps, querying view‑specific features for each 3D point, and introducing a lightweight attenuation‑prediction head. This design yields our first model, Trans‑CBCT, which outperforms the best baseline by 1.17 dB in PSNR and 0.0163 in SSIM on LUNA16 dataset with six projection views. Second, we incorporate a neighbor-aware Point Transformer equipped to further enforce volumetric coherence. This module employs explicit 3D positional encodings and a Neighbor‑Aware Attention module that aggregates information from each point’s k nearest spatial neighbors, improving local consistency and spatial smoothness. The resulting model, named Trans$^2$‑CBCT, achieves an additional improvement of 0.63 dB in PSNR and 0.0117 in SSIM over Trans‑CBCT under the same six‑view setting. Extensive experiments on LUNA16 and ToothFairy datasets, with projection numbers ranging from six to ten, demonstrate that Trans‑CBCT and Trans$^2$‑CBCT consistently outperform all prior methods in term of both PSNR and SSIM. These results highlight the effectiveness of combining hybrid CNN–Transformer features with geometry‑aware point-based reasoning for sparse-view CBCT reconstruction.

\end{abstract}


\section{Introduction}
\label{sec:intro}

\IEEEPARstart{E}{arly} X-ray computed tomography (CT) systems relied on a narrow fan-beam geometry: a one-dimensional detector array collected projections as the X-ray source and patient rotated synchronously, producing a stack of 2D sinograms that were reconstructed slice-by-slice. Decades of progress, spanning filtered backprojection algorithms, iterative statistical reconstruction methods~\cite{gordon1970algebraic,ANDERSEN198481,jiang2003convergence}, and advancements in applying different regularization terms~\cite{RUDIN1992259,sidky2008image}, have steadily improved spatial resolution, noise performance, and scan speed. Cone-beam CT (CBCT) extends this evolution by replacing the fan beam with a 2D cone-shaped X-ray beam and a large flat-panel detector. A single gantry rotation now captures an entire volume instead of an individual slice, enabling faster acquisitions and isotropic voxels that are advantageous for dental, head-and-neck, and image-guided interventional applications. However, achieving diagnostic-quality reconstructions traditionally requires hundreds of projection views, leading to prolonged scan times and increased radiation exposure. These concerns drive the development of sparse-view CBCT reconstruction.

Sparse-view CBCT reconstruction aims to reconstruct accurate three-dimensional volumes from a few projection views (typically six to ten views), thereby reducing the scanning time and lowering patient dose. Learning-based methods have shown promise in addressing this challenge~\cite{lin2023learning,lin2024c2rv,zha2022naf,r2_gaussian,li2024intensity}. NeRF-inspired approaches~\cite{zha2022naf,fang2022snaf} treat CBCT as a continuous attenuation field and optimize scan-specific parameters to minimize the discrepancy between real and synthesized projections. Although these methods are capable of capturing fine details, they demand intensive per-scan optimization and exhibit performance degradation when input views drop below ten. In contrast, data-driven methods~\cite{lin2023learning, lin2024c2rv,liu2024geometry} use deep neural networks, such as UNet~\cite{ronneberger2015u} and ResNet~\cite{he2016deep} to extract features from sparse projections, backprojecting these features into 3D space and refining 3D representations through MLPs or convolutional decoders. For instance, DIF-Net~\cite{lin2023learning} utilizes a UNet encoder-decoder to extract image-level features, which are then processed by MLPs to predict voxel-wise intensities. Similarly, C$^2$RV \cite{lin2024c2rv} employs UNet to learn multi-scale, multi-view features that are combined and aligned in multi-scale 3D volumetric representations. Another method~\cite{liu2024geometry} adopts ResNet34~\cite{he2016deep} as the image encoder, projecting learned features into a 3D volume and reconstructing the final CBCT image using a 3D CNN decoder. 
\IEEEpubidadjcol

Despite their successes, a fundamental limitation of the UNet or ResNet-based architectures is their inherent reliance on local convolutional operations, restricting their ability to effectively capture global context and long-range spatial dependencies. In this work, to overcome this bottleneck, we first replace UNet/ResNet with TransUNet~\cite{chen2021transunet}, a hybrid CNN-Transformer whose convolutions capture fine local details while its self-attention layers provide the global context. For each projection, we extract multi-scale TransUNet features, query view-specific features given 3D points, and fuse the 3D features across all views via max-pooling. The aggregated 3D features are then forwarded to a lightweight attenuation-prediction head, yielding our first model, Trans-CBCT, which already surpasses the baselines, in terms of PSNR and SSIM, and improves performance on both LUNA16~\cite{setio2017validation} and ToothFairy~\cite{cipriano2022deep} datasets. Although Trans-CBCT delivers favorable results, the reconstructed volume can still be spatially more coherent. Hence, we also introduce a neighbor-aware \textit{Point Transformer} to refine 3D features and improve spatial consistency. Sparse-view CBCT reconstruction often struggles with maintaining accurate spatial relationships given sparse projection views. The Point Transformer overcomes this challenge by adding 3D coordinate positional encodings to the 3D point features, which enables the network to reason about the spatial positioning of features more effectively. Moreover, we integrate a KNN-based Neighbor-Aware Attention mechanism within the Point Transformer, which helps in preserving local details and promoting spatial smoothness. This mechanism selectively aggregates information from spatially neighboring points, thereby mitigating aliasing artifacts and ensuring that local anatomical details are not lost. Combining TransUNet's global feature extraction with our neighbor-aware Point Transformer's locality-aware refinement yields a unified framework that delivers state-of-the-art accuracy on sparse-view CBCT reconstruction.


To summarize, the main contributions of this work include the following:
\begin{itemize}
    \item We replace the conventional U-Net or ResNet-based architectures with TransUNet, resulting in our first model Trans-CBCT, which combines convolutional encoding for multi-scale feature embeddings and transformer-based self-attention for long-range context.

    \item We append our Trans-CBCT with a Point Transformer that embeds explicit 3D positional encodings into each sampled point, enabling the network to reason directly about spatial geometry and maintain global volumetric consistency.

    \item We augment the Point Transformer with a Neighbor-Aware Attention module that selectively aggregates information from each point's nearest neighbors, reducing aliasing artifacts and preserving local structures.

    \item By integrating TransUNet with coordinate-aware point reasoning and locality-driven attention, our method Trans$^2$‑CBCT achieves high-fidelity 3D reconstructions from extremely sparse (6-10 views) CBCT data, outperforming the existing baselines in terms of both PSNR and SSIM, and providing state-of-the-art performance.
\end{itemize}


\section{Related Work}
\label{sec:related}

\subsection{Sparse-View CBCT Reconstruction}
In sparse-view CBCT, the goal is to reconstruct volumetric images from significantly fewer X-ray projection angles than conventional scans, reducing radiation dose and scan time. However, angular undersampling leads to severe streak artifacts and loss of fine structures when using traditional analytic reconstruction techniques. This challenge has motivated a broad range of methods, including iterative, deep learning approaches, each seeking to balance volumetric image quality, computational efficiency, and radiation safety.

\noindent \textbf{(i) Traditional Methods.} Feldkamp-Davis-Kress (FDK)~\cite{Feldkamp:84} algorithm applies a weighted 1D filtering followed by filtered backprojection (FBP) along divergent X-ray paths. Despite its speed and simplicity, FDK produces pronounced streak artifacts and resolution loss under sparse-view sampling, because it lacks mechanisms to enforce consistency across views or incorporate prior information. To address these artifacts, iterative reconstruction methods formulate CBCT as an optimization problem, incorporating regularization terms to impose sparsity smoothness.~Algebraic Reconstruction Technique (ART)~\cite{gordon1974tutorial} and Simultaneous Algebraic Reconstruction Technique (SART)~\cite{ANDERSEN198481} iterate between forward and backward projections to progressively refine the image. Total variation (TV) minimization~\cite{sidky2008image} and its variants~\cite{cui2024sparse,Xu:20} enforce sparsity of image gradients, effectively reducing streak artifacts in sparse-view data. 

\noindent \textbf{(ii) Deep Learning-Based Methods.} With the rise of deep neural networks, data-driven methods have achieved remarkable results. Deep Back Projection~\cite{ye2018deep} feeds individual backprojected views into a CNN to learn a mapping from noisy backprojections to artifact-free images. FBPConvNet~\cite{7949028} applies a UNet to FBP-reconstructed images, training on paired sparse-view and full-view reconstruction. Recent implicit neural representation (INR) methods treat the attenuation field as a continuous function parameterized by a coordinate-based MLP. PixelNeRF~\cite{yu2021pixelnerf} and its adaptations~\cite{shen2022nerp,zha2022naf} leverage sparse views to condition neural field on projection data, enabling high-fidelity reconstructions under extreme sparse-view setting. Such INR approaches implicitly encode volumetric consistency and have shown promising results, particularly when combined with Gaussian representations~\cite{lin2024learning, r2_gaussian}. DIF‑Net~\cite{lin2023learning} uses a 2D U‑Net to extract feature maps from each input view, projects arbitrary 3D query points into those maps to sample view‑specific features, fuses them via an MLP, and regresses per‑point intensities. DIF‑Gaussian~\cite{lin2024learning} also uses U‑Net feature extractor but replaces dense voxels with 3D Gaussians to embed spatial priors, improving point‑wise intensity prediction. C$^2$RV~\cite{lin2024c2rv} further extends U‑Net by extracting multi‑scale 2D feature maps, back‑projecting them into explicit 3D volumes for cross‑regional learning, and using a scale‑view cross‑attention module to fuse voxel‑aligned and pixel‑aligned features. While all three methods~\cite{lin2023learning,lin2024c2rv,lin2024learning} leverage U‑Net’s strong local feature encoding, they lack an explicit mechanism for modeling long‑range spatial dependencies or global context across the entire image plane—an important factor when reconstructing from extremely sparse views. To address this, we replace the standard U‑Net encoder with TransUNet~\cite{chen2021transunet}, which integrates a vision-transformer block into the bottleneck. This hybrid design preserves UNet’s high‑resolution, multi‑scale feature maps while adding self‑attention layers that capture non‑local correlations and global structure. By adopting TransUNet, our framework yields richer, more globally coherent image features that translate into more accurate CBCT reconstruction that is robust to artifacts under extreme sparsity.

\subsection{Transformers in Medical Imaging}
Transformer architectures, initially popularized for images by Vision Transformers (ViT)\cite{dosovitskiy2020image}, have been increasingly adopted in medical imaging tasks, including classification and segmentation. Owing to their ability to model long-range dependencies, Transformers have emerged as powerful alternatives to traditional CNNs, especially when capturing complex global contexts is critical.

\noindent \textbf{(i) Transformers for Medical Image Classification.}~Transformer architectures have been applied to different medical image modalities to achieve classification.~TL-Med~\cite{MENG2022842} employs a two-stage transfer learning strategy utilizing the ViT architecture. At first stage, ViT model is pretrained on the ImageNet~\cite{5206848} to capture generic features. Then model is fine-tuned on a tuberculosis CT scan dataset. At the second stage, the model is further fine-tuned on a COVID-19 CT scan dataset. This stage refines the model's ability to distinguish between COVID-19 and non-COVID-19 cases.~\cite{dai2021transmed} introduces Transformer into multi-modal medical image classification by modeling both spatial and modality-wise correlations through self-attention, achieving superior performance in handling missing or noisy modalities. Mohamed et al.~\cite{gheflati2022vision} proposed the use of ViTs for breast ultrasound image classification, demonstrating that the self-attention mechanism enables more effective modeling of long-range dependencies compared to conventional CNNs. By directly processing flattened image patches, the ViT architecture better captures global anatomical patterns, leading to improved classification performance, particularly in distinguishing malignant from benign lesions.

\noindent \textbf{(ii) Transformers for Medical Image Segmentation.}~Transformers have also shown promise in medical image segmentation, where capturing both local details and global anatomical context is crucial. Swin UNETR~\cite{hatamizadeh2021swin} leverages a Swin Transformer~\cite{liu2021swin} as the encoder within a U-shaped architecture for brain tumor segmentation in MRI scans. By using hierarchical feature maps and local window-based self-attention, Swin UNETR significantly improves segmentation accuracy compared to CNN-based methods.~VT-UNet~\cite{peiris2022robust} pushed this further by removing convolutional components entirely and using ViT throughout the architecture.~VT-UNet employs hierarchical volumetric self-attention with shifted windows, parallel self- and cross-attention in the decoder, and Fourier positional encoding, achieving good tumor segmentation accuracy and robustness with lower computational complexity. Meanwhile, specialized designs, such as D-Former~\cite{wu2023d}, introduces a novel self-attention mechanism that alternates between local scope modules and global scope modules, and uses 3D depth-wise convolutions to learn position information dynamically. With these modules, D-Former offers a good trade-off between accuracy and computational cost.~TransUNet\cite{chen2021transunet} combines CNN-based encoders with Transformer modules to simultaneously capture local fine-grained features and global context, significantly improving segmentation performance. We adopt TransUNet as the 2D feature extractor for X-ray projection images due to its architectural simplicity and effectiveness in 2D medical image analysis. Unlike Swin UNETR and VT-UNet, which are primarily designed for 3D volumetric segmentation, TransUNet is more suitable for our task, which operates on 2D projection images. Moreover, its modular CNN-Transformer structure allows us to easily extract and concatenate multi-scale features, which is crucial for our cross-view fusion and 3D point-wise prediction.

\noindent \textbf{(iii) Transformers for CBCT Images.}
Recently, Transformer architecture has also been explored for improving CBCT images in various ways. For example, Cao et al.~\cite{CAO2025108637} proposed a 4D-CBCT method that leverages Swin Transformers~\cite{liu2021swin} and masking strategies to improve temporal consistency and reconstruction robustness. Yuan et al.~\cite{YUAN2025107197} introduced WUTrans, a whole-spectrum unilateral-query-secured Transformer tailored for 4D CBCT reconstruction, aiming to model both spatial and temporal dynamics with improved query efficiency. Zhao et al.~\cite{zhao2024cbctformer} presented CBCTformer, a Transformer-based artifact reduction model designed to enhance already-reconstructed CBCT volumes. Sha et al.~\cite{sha2024removing} further address ring artifact suppression using a Transformer model trained with a dual-domain loss, including a novel unidirectional vertical gradient loss in polar coordinates. Gao et al.~\cite{gao2023transformer} presented Dual-Swin, a dual-domain reconstruction framework that reconstructs field-of-view (FOV) extended CBCT image from truncated sinograms by applying Swin Transformers in both sinogram and image domains. While effective for addressing truncation artifacts in radiation therapy, Dual-Swin assumes relatively dense angular sampling and focuses on sinogram restoration, whereas our work targets direct 3D reconstruction from extremely sparse projections using a unified dual-Transformer framework.

While the aforementioned  works demonstrate the growing role of Transformers in CBCT-related tasks, they often rely on relatively dense projection data or operate in a post-reconstruction refinement setting. In contrast, our work targets the challenging problem of direct 3D CBCT reconstruction from extremely sparse views. We introduce a unified dual-Transformer framework that adapts TransUNet for 2D projection-level feature extraction and introduces a neighbor-aware Point Transformer for 3D spatial reasoning, enabling high-fidelity volumetric reconstruction from as few as six X-ray projections.


\section{Proposed Method}
\label{sec:method}

\subsection{Revisit of TransUNet}

\begin{figure}[t]
  \centering
   \includegraphics[width=\linewidth]{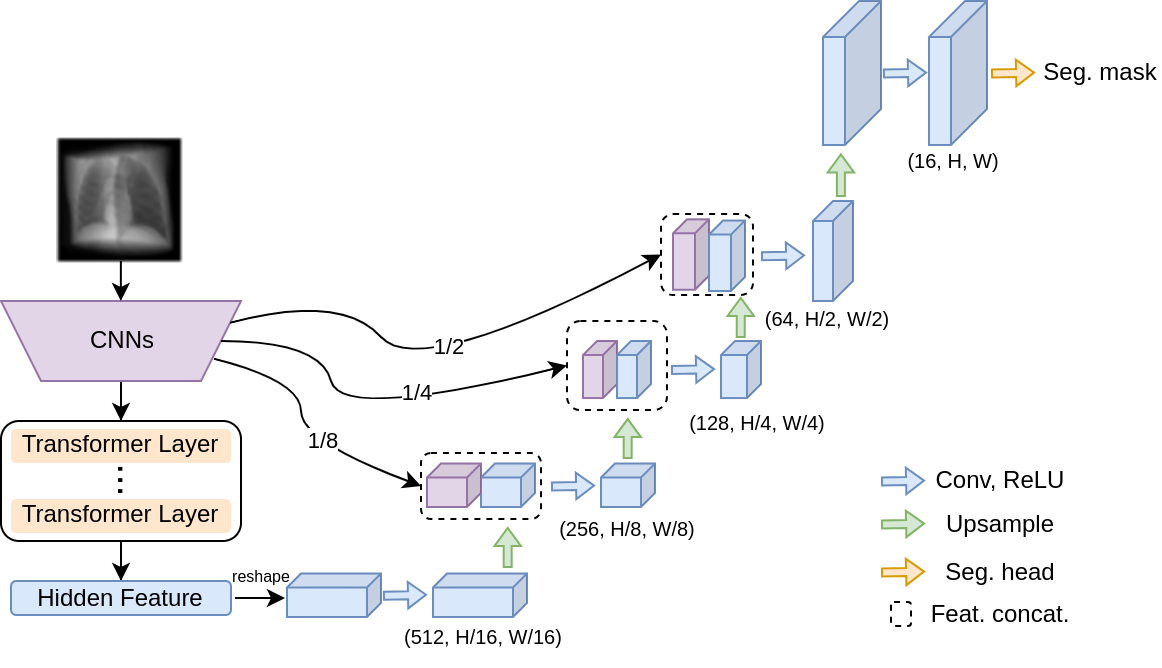}

   \caption{The architecture of TransUNet. It includes a CNN-Transformer hybrid model as encoder and a cascaded upsampler to decode the hidden features.}
   \label{fig:transunet}
\end{figure}

TransUNet~\cite{chen2021transunet} significantly advanced medical image segmentation by combining the complementary strengths of convolutional neural networks (CNNs) and vision transformers. As illustrated in Fig.~\ref{fig:transunet}, TransUNet employs a hybrid architecture, where an initial convolutional encoder captures hierarchical local features from the input images. These CNN-derived features are then processed by transformer layers, utilizing self-attention mechanisms to effectively capture long-range, global contextual relationships.

A notable innovation of TransUNet lies in its U-shaped structure, which integrates globally-aware transformer representations with detailed CNN features through skip connections and upsampling operations. This design ensures that crucial fine-grained details are preserved while simultaneously benefiting from the transformer's global context modeling, thereby enhancing segmentation accuracy. With the u-shaped CNN-Transformer hybrid design, TransUNet demonstrates superior segmentation performance compared to purely convolutional or transformer-based methods.

Given these advantages, we hypothesize that TransUNet’s robust feature extraction capabilities, specifically its proficiency in simultaneously modeling detailed local information and global dependencies, can extend effectively beyond segmentation tasks. In this work, we show that TransUNet's architecture is well-suited for sparse-view CBCT reconstruction, where precise and globally-aware feature extraction is critical to reconstruct high-quality volumetric images from extremely limited X-ray projection data.

\subsection{The Architecture of Proposed Trans-CBCT}
\begin{figure*}[t]
  \centering
   \includegraphics[width=\linewidth]{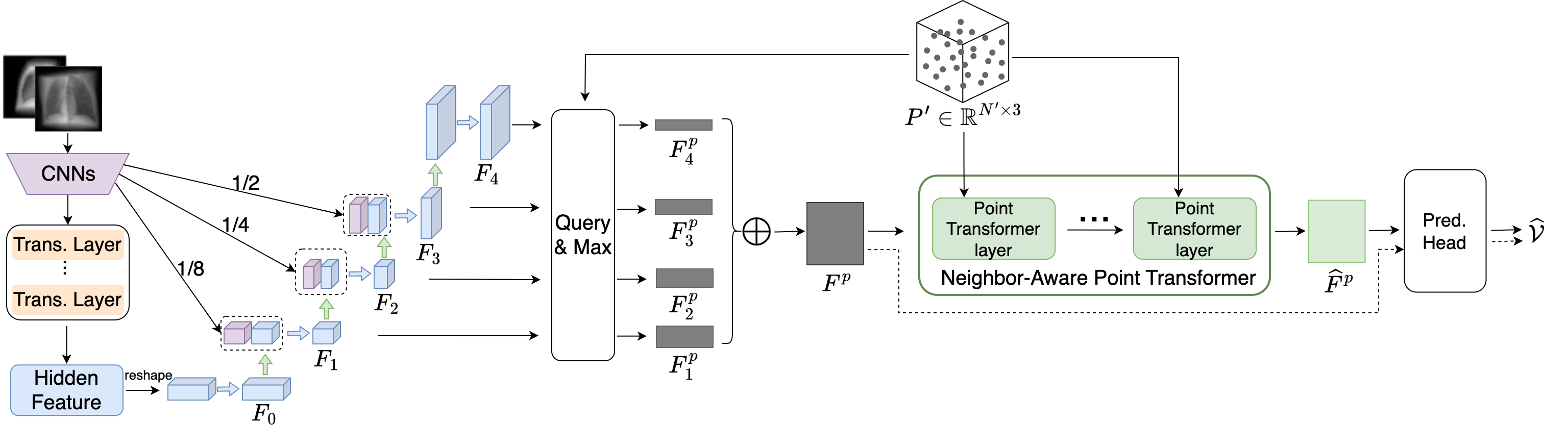}

   \caption{The architecture of Trans-CBCT and Trans$^2$-CBCT. Projection images are first encoded using TransUNet to extract multi-scale feature maps. Features from multiple decoder levels are queried and concatenated ($\oplus$) into point features $F^p$. Feeding $F^p$ directly to the prediction head (dashed arrow) produces the Trans-CBCT model. Alternatively, passing $F^p$ through a Point Transformer yields spatially-aware features $\widehat{F}^p$, which are then used for final prediction in the Trans$^2$-CBCT model.}
   \label{fig:framework}
\end{figure*}

Following prior works~\cite{lin2024c2rv,lin2023learning,zha2022naf,lin2024learning}, we formulate the CBCT reconstruction as learning a continuous intensity field. Given a 3D spatial domain $\mathcal{P} $ and $M$ sparse projection images $\mathcal{I}=\{I_1,...,I_M\}$, the goal is to learn a continuous function $f$ mapping each spatial point $p \in \mathcal{P}$ to its attenuation coefficient:
\begin{equation}
    \mathcal{\widehat{V}}=\{\hat{v}(p)=f(\mathcal{I},p)\,\,|\,\,\forall p \in \mathcal{P}\}.
\end{equation}
To effectively extract meaningful feature representations from limited projections, we employ TransUNet as a feature encoder, shared across all $M$ projections. Specifically, TransUNet extracts multi-scale feature maps $\{F_{s,1},...,F_{s,M}\}$, where each feature map $F_{s,m} \in \mathbb{R}^{W_s \times H_s \times C_s}$ corresponds to scale $s \in {1,2,3,4}$ for view $m$, with $W_s$, $H_s$, and $C_s$ denoting the width, height and number of channels of the feature map at scale $s$, respectively. As illustrated in Fig.~\ref{fig:framework} (the subscript $m$ is omitted in the figure for clarity), we aggregate features from multiple stages of the decoder rather than relying solely on the final-level features. This multi-scale fusion strategy ensures that our model simultaneously captures both low-level spatial features and high-level semantic context.

During training, for a given point $p \in \mathcal{P}$, we first project $p$ onto each of the $M$ image planes using the corresponding projection function $\pi_m$. We then compute view-specific features by performing bilinear interpolation on the feature maps at each scale:
\begin{equation}
\small
    F^p_{s,m}(p) = \text{Interp}(F_{s,m},\pi_m(p)) \in \mathbb{R}^{C_s}.
     \label{eq:interp}
\end{equation}
where $Interp(\cdot)$ is the bilinear interpolation operator.

Next, we fuse these view-specific features across all $M$ projections using max-pooling:
\begin{equation}
\small
    F^p_{s}(p) = \text{max}(\{F^p_{s,1}(p), F^p_{s,2}(p),...,F^p_{s,M}(p)\}) \in \mathbb{R}^{C_s}.
    \label{eq:max}
\end{equation}

Finally, we concatenate ($\oplus$) the fused features across different scales, $s \in {1,2,3,4}$, to produce a rich set of multi-scale features for point $p$:
\begin{equation}
\small
    {F}^p(p) =  {F}^p_1(p) \oplus {F}^p_2(p) \oplus {F}^p_3(p) \oplus {F}^p_4(p) \in \mathbb{R}^{256+128+64+16}
     \label{eq:cat}
\end{equation}

\subsection{The Architecture of Proposed Trans$^2$-CBCT}
We enhance our Trans-CBCT framework by incorporating our neighbor-aware Point Transformer module, resulting in our full framework, referred to as the Trans$^2$-CBCT.

\textbf{Neighbor-Aware Point Transformer Layer.} 
While view-specific features are essential for reconstructing the 3D volume from X-ray projection images, enforcing spatial consistency among voxels is equally important for producing artifact-suppressed and anatomically plausible reconstructions. To promote local smoothness and geometric coherence, we introduce a neighbor-aware Point Transformer module.

\begin{figure}[t]
  \centering
   \includegraphics[width=\linewidth]{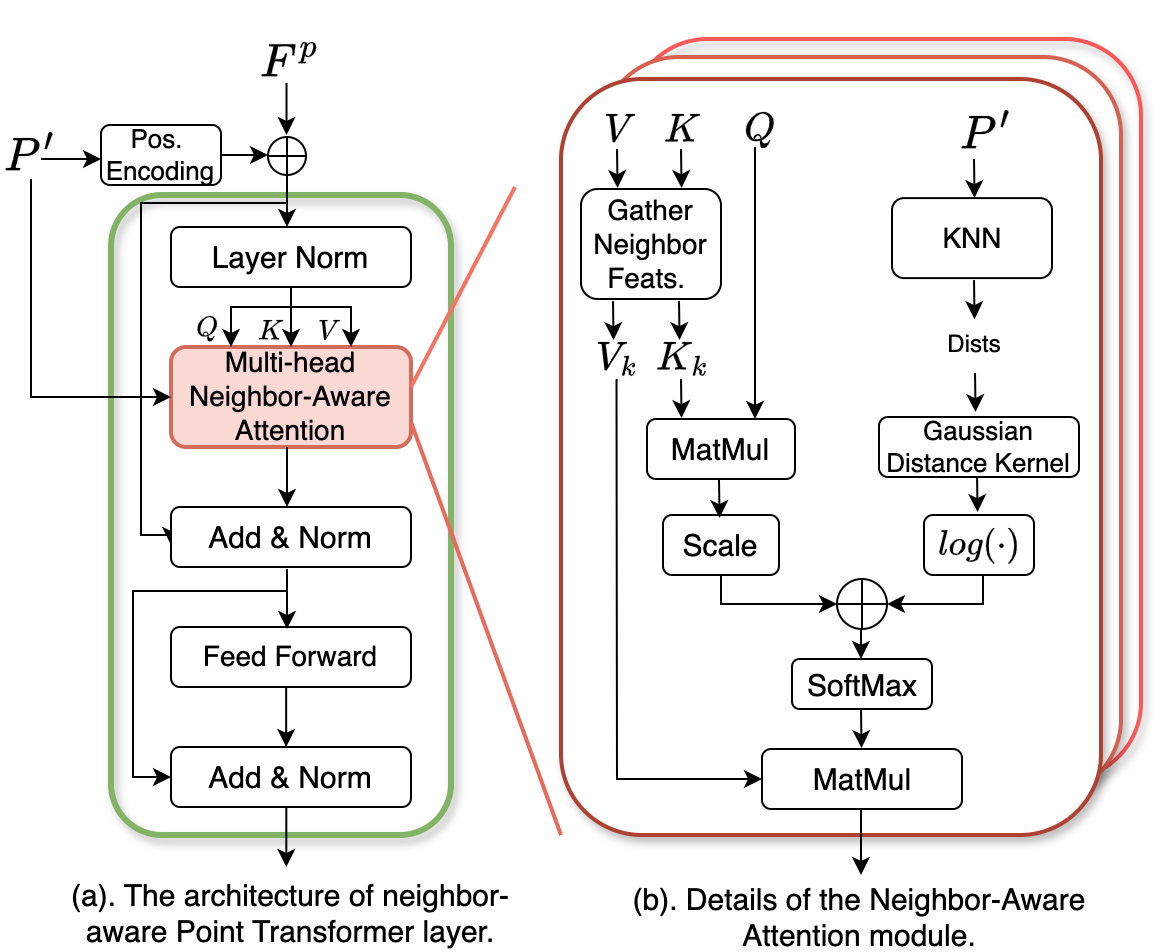}

   \caption{The architecture of neighbor-aware Point Transformer layer. (a) Each sampled 3D point is embedded with positional encoding and processed through transformer layers; (b) Multi-head neighbor-aware attention is restricted to $k$-nearest neighbors, with attention scores guided by both feature similarity and a Gaussian-weighted distance bias, improving local smoothness and geometric consistency.}
   \label{fig:pointtrans}
\end{figure}

Rather than processing every 3D point individually, we randomly sample $N'$ points from the 3D space $\mathcal{P}$, yielding a set $P' \in \mathbb{R}^{N' \times 3}$. For each sampled point, we query view-specific features using 
 Eqs.~(\ref{eq:interp}),~(\ref{eq:max}), and~(\ref{eq:cat}), obtaining a fused multi-scale feature matrix ${F}^p \in \mathbb{R}^{N' \times 464}$.
As illustrated in Fig.~\ref{fig:pointtrans}(a), the input to the Point Transformer consists of the point coordinates $P'$ and their associated features $\mathcal{F}^p$. To make the transformer aware of 3D geometry, we embed each point $p_i \in P'$ using a learnable positional encoding:
\begin{equation}
\small
    PE(p_i) = \text{max}(0, p_iW_1^T+b_1)W_2^T+b_2,
\end{equation}
where $W_1$, $W_2$ are learnable weight matrices, and $b_1$, $b_2$ are bias terms. The positional encoding is added to the multi-scale features ${F}^p$ before feeding them into our neighbor-aware Point Transformer.

\textbf{Multi-Head Neighbor-Aware Attention.} 
Traditional self-attention mechanisms compute pairwise relationships among all input tokens, which can be computationally expensive and may overlook valuable local structures in spatial data. To encourage local smoothness and reduce computational complexity, we propose a multi-head neighbor-aware attention mechanism, illustrated in Fig.~\ref{fig:pointtrans}(b). For each point $p_i \in P'$, we identify its $k$-nearest neighbors using a KNN algorithm. From the Euclidean distance between $p_i$ and a neighbor $p_j$, denoted by $d_{ij}$, we compute a Gaussian-weighted adjacency:
\begin{equation}
\small
    w_{ij} = \exp\left(-\frac{d_{ij}^2}{2\sigma^2}\right),
\end{equation}
where $\sigma$ controls the sharpness of the decay. This weighting ensures that nearby points make stronger influence during aggregation.

Each attention head linearly transforms input features into queries $Q$, keys $K$, and values $V$. However, rather than attending to all points, we restrict attention to each point's local neighborhood. We compute similarity scores as:
\begin{equation}
\small
    \text{score}_{ij} = \frac{Q_i\cdot K_j^T}{\sqrt{d_k}} + log(w_{ij}),
\end{equation}
where $d_k$ is the dimension of key vectors. The scores are normalized using a softmax function:
\begin{equation}
\small
    \alpha_{ij} = \text{SoftMax}(\text{score}_{ij}).
\end{equation}
The output for each point is then computed by a weighted sum of its neighbors' value features:
\begin{equation}
\small
\widehat{F}^p(p_i) = \sum_{j \in \mathcal{N}(i)} \alpha_{ij} V_j,
\end{equation}
where $\mathcal{N}(i)$ denotes the set of $k$ nearest neighbors of point $p_i$.
By employing multiple attention heads, we allow the network to capture diverse local patterns. Consequently, our neighbor-aware attention combines both spatial distance and feature similarity to produce robust, context-rich features.

Our neighbor-aware Point Transformer module consists of $l$ such layers stacked sequentially. By incorporating geometry-aware positional encodings and neighbor-aware attention, our neighbor-aware Point Transformer promotes local smoothness and spatial consistency in the volumetric representation. This design significantly enhances reconstruction quality by explicitly modeling the spatial structure of the volume, particularly under conditions of extreme projection sparsity.

\subsection{Model Optimization}
For both of our proposed frameworks, Trans-CBCT and Trans$^2$-CBCT, the feature representation $\widehat{F}^p(p_i)$ for each point $p_i$ is passed through a lightweight attenuation prediction head. This head consists of a sequential stack of Conv1D, BatchNorm, ReLU, Conv1D, and Sigmoid layers, which outputs the predicted attenuation coefficient $\hat{v}_i$.

Rather than directly reconstructing the entire 3D CT volume, we adopt a point-wise regression strategy. Specifically, we sample a set of points $P'$ from the ground-truth CT volume and train the network to predict the corresponding attenuation coefficients. The ground-truth $\mathcal{V} = \{v_1, v_2,...,v_{N'}\}$ for these sample points are obtained by trilinear interpolation from the full-resolution CT volume.

We optimize the model using the mean square error (MSE) loss between the predicted and ground-truth coefficients:
\begin{equation}
\small
    \mathcal{L}(\mathcal{V}, \mathcal{\widehat{V}}) = \frac{1}{N'} \sum_{i=1}^{N'} (v_i-\hat{v}_i)^2.
\end{equation}

\section{Experiments}
\label{sec:exp}

\begin{table*}[tbp]
\caption{Quantitative comparison of baselines on two datasets given 6/8/10 projection views. Higher values of PSNR (dB) and SSIM ($\times 10^{-2}$) indicate better reconstruction quality. Our models Trans$^2$-CBCT and Trans-CBCT achieve the best and second-best performance on both datasets, shown as \textbf{bold} and \underline{underlined}, respectively.}
\resizebox{\linewidth}{!}{
\begin{tabular}{c|c|cccccc|cccccc}
\hline
\multicolumn{1}{c|}{\multirow{3}{*}{Method}} & \multicolumn{1}{c|}{\multirow{3}{*}{Type}}                                                               & \multicolumn{6}{c|}{LUNA16}                                                           & \multicolumn{6}{c}{ToothFairy}                                                        \\ \cline{3-14} 
                        &                                                                                   & \multicolumn{2}{c}{6-View} & \multicolumn{2}{c}{8-View} & \multicolumn{2}{c|}{10-View} & \multicolumn{2}{c}{6-View} & \multicolumn{2}{c}{8-View} & \multicolumn{2}{c}{10-View} \\ 
  &                                                        & \multicolumn{1}{c}{PSNR} & \multicolumn{1}{c}{SSIM} & \multicolumn{1}{c}{PSNR} & \multicolumn{1}{c}{SSIM} & \multicolumn{1}{c}{PSNR} & \multicolumn{1}{c|}{SSIM} & \multicolumn{1}{c}{PSNR} & \multicolumn{1}{c}{SSIM} & \multicolumn{1}{c}{PSNR} & \multicolumn{1}{c}{SSIM} & \multicolumn{1}{c}{PSNR} & \multicolumn{1}{c}{SSIM} \\ \hline      \hline                  
FDK~\cite{Feldkamp:84}                     & \multirow{4}{*}{Self-Supervised}                                                  & 15.34        & 35.78       & 16.58        & 37.89       & 17.40        & 39.85        & 17.07        & 39.90       & 18.42        & 43.29       & 19.58        & 47.21        \\ 
SART~\cite{ANDERSEN198481}                    &                                                                                   & 19.70        & 64.36       & 20.06        & 67.80       & 20.23        & 70.23        & 20.04        & 64.98       & 21.92        & 67.86       & 22.82        & 71.53        \\ 
NAF~\cite{zha2022naf}                     &                                                                                   & 18.76        & 54.16       & 20.51        & 60.84       & 22.17        & 62.22        & 20.58        & 63.52       & 22.39        & 67.24       & 23.83        & 72.52        \\ 
NeRP~\cite{shen2022nerp}                    &                                                                                   & 23.55        & 74.46       & 25.83        & 80.67       & 26.12        & 81.30        & 21.77        & 72.06       & 24.18        & 78.83       & 25.99        & 82.08        \\ \hline
FBPConvNet~\cite{7949028}              & \multirow{3}{*}{\begin{tabular}[c]{@{}c@{}}Data-Driven:\\ Denoising\end{tabular}} & 24.38        & 77.57       & 24.87        & 78.86       & 25.90        & 80.03        & 27.22        & 79.33       & 27.72        & 81.90       & 28.13        & 83.51        \\ 
FreeSeed~\cite{ma2023freeseed}                &                                                                                   & 25.59        & 77.36       & 26.86        & 78.92       & 27.23        & 79.25        & 26.35        & 78.98       & 27.08        & 81.38       & 27.63        & 84.40        \\ 
BBDM~\cite{li2023bbdm}                    &                                                                                   & 24.78        & 77.03       & 25.81        & 78.06       & 26.35        & 79.38        & 26.29        & 78.57       & 27.28        & 80.33       & 28.00        & 83.96        \\ \hline
PixelNeRF~\cite{yu2021pixelnerf}               & \multirow{6}{*}{\begin{tabular}[c]{@{}c@{}}Data-Driven:\\ INR-based\end{tabular}} & 24.66        & 78.68       & 25.04        & 80.57       & 25.39        & 82.13        & 24.85        & 80.91       & 25.37        & 82.11       & 25.90        & 83.25        \\  
DIF-Net~\cite{lin2023learning}                 &                                                                                   & 25.55        & 84.40       & 26.09        & 85.07       & 26.67        & 86.09        & 25.78        & 83.62       & 26.29        & 84.81       & 26.90        & 86.42        \\ 
DIF-Gaussian$^*$~\cite{lin2024learning}            &                                                                                   & 27.71        & 84.87       & 28.71        & 86.43       & 29.07             & 87.31             & 28.81             & 87.06            & 29.83             & 88.67            & 29.07             &  85.17            \\ 
C$^2$RV~\cite{lin2024c2rv}                    &                                                                                   & 29.23        & 87.47       & 29.95        & 88.46       & 30.70        & 89.16        & -            & -           & -            & -           & -            & -            \\ 
Trans-CBCT (ours)                    &                                                                                   & \underline{30.40}         & \underline{89.10}      & \underline{31.12}        & \underline{90.07}       & \underline{31.82}        & \underline{90.88}        &  \underline{31.04}            & \underline{91.09}            & \underline{32.11}             & \underline{92.32}            & \underline{32.55}             & \underline{92.72}             \\ 
Trans$^2$-CBCT (ours)                    &                                                                                   & \textbf{31.03}       & \textbf{90.27}       & \textbf{31.70}        & \textbf{91.05}       & \textbf{32.44}        & \textbf{91.88}        & \textbf{31.60}         & \textbf{92.16}        & \textbf{32.71}        & \textbf{93.27}       & \textbf{33.63}        & \textbf{93.96}        \\ \hline

\end{tabular}
}
\label{tab:exp}
\end{table*}

\subsection{Experimental Details}

\textbf{Datasets.}
We conduct experiments on two public datasets: LUNA16~\cite{setio2017validation}, which is a chest CT dataset, and ToothFairy~\cite{cipriano2022deep}, which is a dental CBCT dataset. Following previous works~\cite{lin2023learning, lin2024learning}, we split LUNA16 into 738 scans for training, 50 scans for validation and 100 scans for testing. The chest CT scans are preprocessed so that each volume has shape $256\times 256 \times 256$ with consistent spacing $[1.6, 1.6, 1.6]$ mm. ToothFairy is divided into 343 scans for training, 25 scans for validation and 75 scans for testing. The dental CBCT scans have volume shape $256 \times 256 \times 256$ with spacing $[2.1, 5.4, 5.4]$ mm. We use digitally reconstructed radiographs (DRRs) to generate the projection images with resolution of $256 \times 256$, then randomly sample projection images within the range of $[0\degree, 180\degree)$.

\textbf{Experimental Setup.}
All the experiments are conducted on an NVIDIA RTX 6000. The model is optimized with AdamW~\cite{loshchilov2017decoupled} optimizer with a learning rate of 0.001 and weight decay of $1 \times 10^{-6}$. The model is trained for 400 epochs with batch size of 2 for both datasets. $N'=10,000$ points are sampled as part of the input. To balance computational efficiency and reconstruction accuracy, we use $l=2$ Point Transformer layers and set $k=3$ for KNN algorithm in Neighbor-Aware Attention. We use peak signal-to-noise ratio (PSNR) and structural similarity index measure (SSIM) as the performance metrics. Higher values mean better reconstruction results.

\textbf{Baselines.}
Following~\cite{lin2024learning}, we compare our model with three types of baselines, namely self-supervised methods, denoising-based data-driven methods and implicit neural representation (INR)-based data-driven methods. Self-supervised methods, which require no external CT scans but only projection images, include FDK~\cite{Feldkamp:84}, SART~\cite{ANDERSEN198481}, NAF~\cite{zha2022naf} and NeRP~\cite{shen2022nerp}. Denoising-based methods include FBPConvNet~\cite{7949028}, FreeSeed~\cite{ma2023freeseed} and BBDM~\cite{li2023bbdm}. INR-based methods include PixelNeRF~\cite{yu2021pixelnerf}, DIF-Net~\cite{lin2023learning}, DIF-Gaussian~\cite{lin2024learning} and C$^2$RV~\cite{lin2024c2rv}. For DIF-Gaussian$^*$~\cite{lin2024learning} in Tab.~\ref{tab:exp}, we reproduced the experiments given the original authors' implementation. The remaining results are taken from DIF-Gaussian~\cite{lin2024learning}.

\subsection{Quantitative Results}
We evaluate all methods at a fixed resolution of $256^3$ using 6, 8, and 10 projection views. As shown in Table~\ref{tab:exp}, self‑supervised reconstructions (FDK, SART, NeRP) yield the lowest fidelity, caused by lacking learned priors to suppress artifacts. Denoising networks (FBPConvNet, BBDM) improve results by applying data‑driven post‑processing. Implicit neural representation-based methods (PixelNeRF, DIF‑Net, DIF‑Gaussian, C$^2$RV) further advance performance by modeling the continuous attenuation field and incorporating learned priors. Notably, on LUNA16 with just six views, our Trans‑CBCT framework already exceeds the best baseline (INR‑based C$^2$RV) by $1.17$ dB in PSNR and $1.63 \times 10^{-2}$ in SSIM, demonstrating the enhanced representational capacity of TransUNet’s hierarchical feature maps. Building on this, Trans$^2$‑CBCT, with its Point Transformer and neighbor‑aware attention, delivers additional improvement over Trans‑CBCT under the same six‑view setting, and provides an increase of $1.8$ dB in PSNR and $ 2.8 \times 10^{-2}$ in SSIM over the best baseline (INR‑based C$^2$RV).
A similar trend holds on the ToothFairy dataset, where Trans-CBCT and Trans$^2$-CBCT consistently outperform all baselines across all settings. At the 6-view setting, Trans-CBCT surpasses DIF-Gaussian by $2.23$ dB in PSNR and $4.03 \times 10^{-2}$ in SSIM, while Trans$^2$-CBCT further improves performance and surpasses DIF-Gaussian by $2.79$ dB in PSNR and $5.1 \times 10^{-2}$ in SSIM. 
These results confirm that explicit 3D spatial reasoning and local smoothness regularization are pivotal for high‑fidelity sparse‑view CBCT reconstruction.

\subsection{Qualitative Results}
\begin{figure*}[t]
  \centering
   \includegraphics[width=\linewidth]{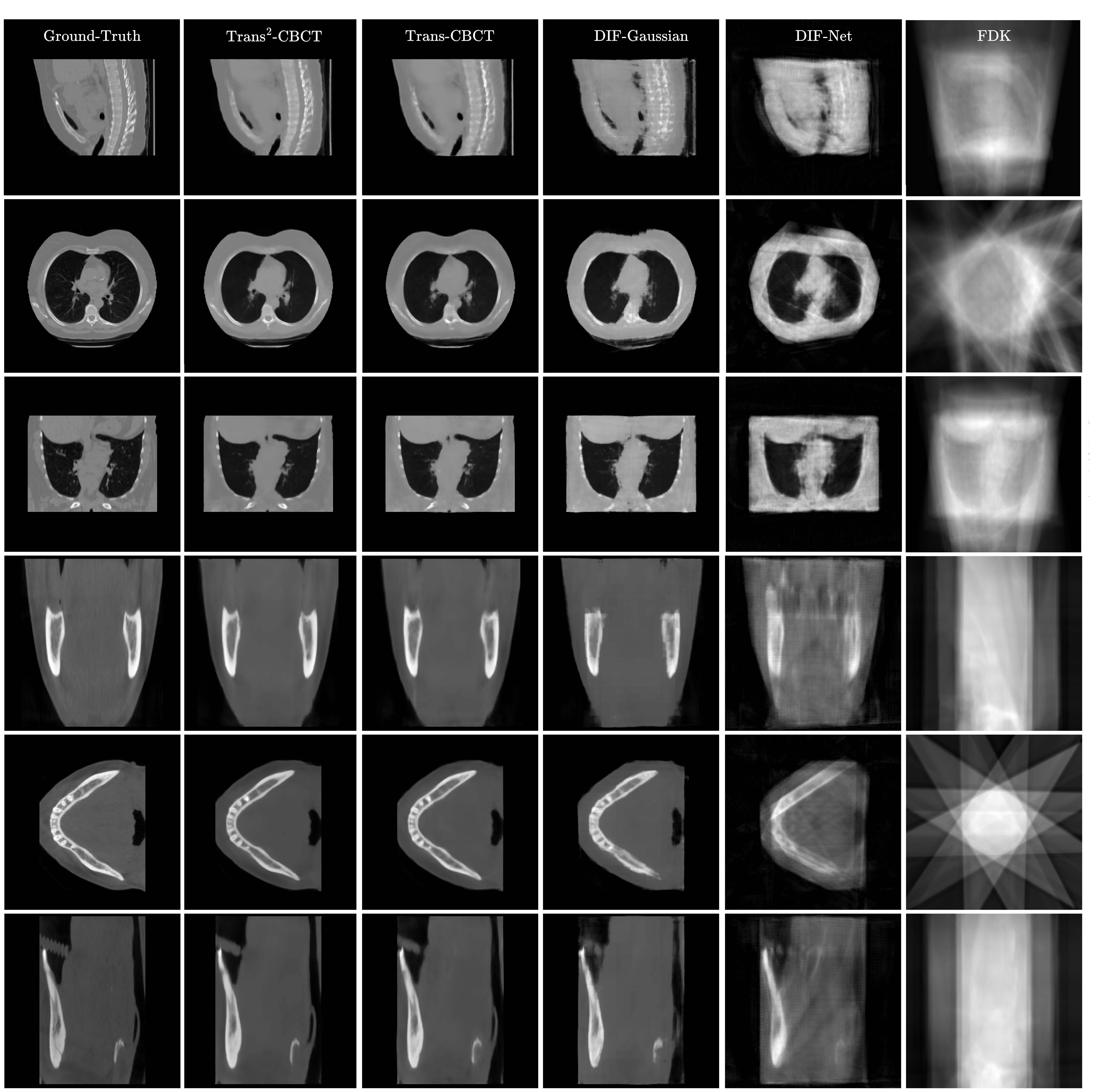}

   \caption{Visualization of 6-view reconstruction results for different methods. Top three rows: axial, coronal, and sagittal slices from the LUNA16 dataset. Bottom three rows: corresponding slices from the ToothFairy dataset.}
   \label{fig:visual}
\end{figure*}

We provide a qualitative comparison of 6-view reconstruction results across different methods in Fig.~\ref{fig:visual}. Overall, the differences among methods highlight the challenges of extremely sparse-view CBCT reconstruction and the advantages of our proposed approaches.

The FDK algorithm~\cite{Feldkamp:84} struggles under such sparse-view settings, and cannot reconstruct meaningful anatomical structures, resulting in highly blurred images dominated by streaking and shading artifacts.

DIF-Net~\cite{lin2023learning} shows clear improvements over FDK. It successfully recovers the overall anatomical contours, such as the lung boundaries and the dental arch shapes. However, it still suffers from noticeable limitations: many fine structures are lost, and strong streak artifacts remain present throughout the volume, particularly around high-contrast edges. This suggests that DIF-Net has difficulty capturing intricate details from extremely limited projections.

DIF-Gaussian~\cite{lin2024learning} further improves upon DIF-Net by introducing a more expressive Gaussian parameterization. As shown in the forth column of Fig.~\ref{fig:visual}, DIF-Gaussian produces cleaner reconstructions with sharper boundaries and better preservation of internal structures. Nevertheless, some blurring and smoothing effects are still evident, especially in regions requiring high-frequency detail recovery, such as small vessel structures in chest images or fine bone textures in dental scans.

Our proposed Trans-CBCT and Trans$^2$-CBCT models achieve the best visual quality across both the chest (LUNA16) and dental (ToothFairy) datasets. Specifically, Trans-CBCT generates highly detailed reconstructions with sharp tissue interfaces and significantly reduced streak artifacts compared to previous baselines. Building on this, Trans$^2$-CBCT further refines local anatomical details, offering even clearer internal structures and more consistent intensity distributions across slices. 

The performance gains can be attributed to the combination of multi-scale feature extraction by TransUNet and spatially aware local refinement by the neighbor-aware Point Transformer. Together, they enable our model to better integrate sparse-view information while preserving both global structure and local detail. These results clearly demonstrate the effectiveness of our framework for overcoming the severe information loss caused by extremely sparse projection sampling.

\subsection{Efficiency Analysis}
We evaluate the computational efficiency of our proposed models, Trans-CBCT and Trans$^2$-CBCT, against two INR-based baselines, DIF-Net and DIF-Gaussian, with results summarized in Table~\ref{tab:efficency}. In terms of trainable parameters, our models are indeed more complex due to the use of TransUNet, which incorporates Transformer layers known to be parameter-intensive. However, this increased capacity enables stronger feature extraction and better reconstruction performance from sparse-view inputs. Notably, despite the larger model size, Trans-CBCT maintains competitive runtime. It requires only marginally more time (1.5 s) than DIF-Net and outperforms DIF-Gaussian in inference speed (running 1.9 s faster). This suggests that our model architecture is not only effective but also efficiently designed for practical use, particularly given the improved reconstruction quality it delivers. Trans$^2$-CBCT takes 53.7 seconds to reconstruct one case. This overhead stems mainly from the KNN-based neighbor search required by the Neighbor-Aware Attention module, which computes the $k$-nearest neighbors for each of 10,000 sampled points. Although this step improves local smoothness and spatial consistency, it introduces additional computational cost. We identify this as a key area for optimization; implementing a more efficient or approximate KNN algorithm could significantly reduce runtime without sacrificing quality.

Overall, Trans-CBCT offers a compelling balance between reconstruction fidelity and inference efficiency. It demonstrates that Transformer-based architectures, when carefully integrated, can outperform traditional convolutional or MLP-based models not only in accuracy but also in practical runtime.

\begin{table}[tbp]
\centering
\caption{Efficiency analysis comparing trainable parameters and inference time of different methods.}
\begin{tabular}{l|l|l}
\hline
Method         & Trainable Param. (MB) & Inference Time (s) \\ \hline
DIF-Net        & 31.1        & 8.4                \\ \hline
DIF-Gaussian   & 31.7        & 11.8               \\ \hline
Trans-CBCT     & 105.4       & 9.9                \\ \hline
Trans$^2$-CBCT & 108.1       & 53.7               \\ \hline
\end{tabular}
\label{tab:efficency}
\end{table}

\subsection{Ablation Studies}

\textbf{Analysis of the Number of Sampled Points $N'$.}
Tab~\ref{tab:Num_points} shows the effect of the number of sampled points, $N'$, on the performance.
When using $N'=5,000$ points, our Trans$^2$-CBCT framework provides relatively lower PSNR and SSIM values. Increasing $N'$ to $10,000$ significantly boosts performance on both datasets, indicating that denser sampling helps the model learn a more accurate and smoother attenuation field. However, further increasing $N'$ to $20,000$ does not lead to continued improvement. On the contrary, performance slightly degrades compared to $10,000$, suggesting that too many sampled points may introduce redundant information, cause overfitting to sampled training points, or slow down convergence. These results demonstrate that setting $N'=10,000$ reaches the best balance between reconstruction accuracy and training efficiency, and is therefore adopted as the default configuration in our experiments.

\begin{table}[h!]
\centering
\caption{Ablation study on the number of sampled points $N'$ during training.}
\begin{tabular}{c|cc|cc}
\hline
\multirow{2}{*}{N'} & \multicolumn{2}{c|}{LUNA16}      & \multicolumn{2}{c}{ToothFairy}   \\ \cline{2-5} 
                    & \multicolumn{1}{c|}{PSNR} & SSIM & \multicolumn{1}{c|}{PSNR} & SSIM \\ \hline
5,000                & \multicolumn{1}{c|}{28.49}     & 87.13     & \multicolumn{1}{c|}{31.15}     & 91.42     \\ \hline
10,000              & \multicolumn{1}{c|}{31.03}     &   90.27   & \multicolumn{1}{c|}{31.6}     &    92.16  \\ \hline
20,000              & \multicolumn{1}{c|}{30.40}     & 89.71     & \multicolumn{1}{c|}{31.27}     & 91.62     \\ \hline
\end{tabular}
\label{tab:Num_points}
\end{table}

\textbf{Analysis of the Number of Neighbors.}
We conduct an ablation study on the number of nearest neighbors $k$ used in the Neighbor-Aware Attention module, with results summarized in Tab~\ref{tab:k_neighbors}. The experiments are performed on both LUNA16 and ToothFairy datasets. Using $k=3$ achieves the highest reconstruction quality, which indicates that a small number of neighbors is sufficient to capture local geometric structure without over-smoothing fine details. As $k$ increases to 6, 9, and 15, we observe a slight decline in both PSNR and SSIM values. These results suggest that involving too many neighbors in the attention operation causes excessive feature smoothing and blurs fine anatomical boundaries. In addition to better reconstruction quality, using a smaller $k$ also improves computation efficiency. A smaller number of neighbors reduces the complexity of the KNN search and the subsequent attention computation, leading to faster inference and lower memory usage during both training and testing. Therefore, setting $k=3$ offers the best trade-off between reconstruction fidelity and computational efficiency, and is adopted as the default configuration in our final Trans$^2$-CBCT model.
\begin{table}[h!]
\centering
\caption{Ablation study on the number of nearest neighbors $k$ in the Neighbor-Aware Attention.}
\begin{tabular}{l|ll|ll}
\hline
     & \multicolumn{2}{l|}{LUNA16}        & \multicolumn{2}{l}{ToothFairy}    \\ \hline
     & \multicolumn{1}{l|}{PSNR}  & SSIM  & \multicolumn{1}{l|}{PSNR} & SSIM  \\ \hline
k=3  & \multicolumn{1}{l|}{31.03} & 90.27 & \multicolumn{1}{l|}{31.60} & 92.16 \\ \hline
k=6  & \multicolumn{1}{l|}{30.23}      &  89.37     & \multicolumn{1}{l|}{31.50}     & 91.95      \\ \hline
k=9  & \multicolumn{1}{l|}{29.91}      & 89.04      & \multicolumn{1}{l|}{31.40}     & 91.75      \\ \hline
k=15 & \multicolumn{1}{l|}{29.76}      &  88.83     & \multicolumn{1}{l|}{31.09}     & 91.20     \\ \hline
\end{tabular}
\label{tab:k_neighbors}
\end{table}

\textbf{Multi-Level Features.}
We conduct an ablation study on different feature aggregation strategies, as summarized in Tab~\ref{tab:feature_strategy}. Using only feature $F_4$ leads to the lowest reconstruction performance, while adding $F_3$ improves results and adding another intermediate feature $F_2$ further enhances performance. This indicates that integrating mid-level semantic information helps refine fine anatomical structures. The full multi-level feature concatenation, which incorporates $F_1$, $F_2$, $F_3$, and $F_4$, yields the best performance. These results demonstrate that aggregating multi-scale features across multiple decoder stages improves reconstruction fidelity compared to using only deep or partial features. This validates the effectiveness of our multi-level feature aggregation design in capturing both fine details and global structure for sparse-view CBCT reconstruction.

\begin{table}[tbp]
\centering
\caption{Experiments on different feature aggregation strategies. Multi-level feature concatenation significantly improves reconstruction performance.}
\begin{tabular}{l|ll|ll}
\hline
\multirow{2}{*}{Feature Strategy} & \multicolumn{2}{l|}{Trans-CBCT}  & \multicolumn{2}{l}{Trans$^2$-CBCT} \\ \cline{2-5} 
                                  & \multicolumn{1}{l|}{PSNR} & SSIM & \multicolumn{1}{l|}{PSNR}          & SSIM         \\ \hline
F4                                & \multicolumn{1}{l|}{26.82}     & 80.46     & \multicolumn{1}{l|}{28.28}              & 85.32             \\ \hline
F4;F3                             & \multicolumn{1}{l|}{28.05}     & 82.61     & \multicolumn{1}{l|}{29.60}              & 87.81             \\ \hline
F4;F3;F2                          & \multicolumn{1}{l|}{28.48}     &  83.69    & \multicolumn{1}{l|}{30.18}              & 88.89             \\ \hline
Full multi-level                  & \multicolumn{1}{l|}{30.40} & 89.10 & \multicolumn{1}{l|}{31.03}         & 90.27        \\ \hline
\end{tabular}
\label{tab:feature_strategy}
\end{table}
\section{Conclusion}
\label{sec:conclusion}

In this paper, we have proposed Trans-CBCT and Trans$^2$-CBCT, two transformer-based frameworks, for reconstructing high-quality CBCT volumes from extremely sparse X-ray projection views. Trans-CBCT employs TransUNet to extract multi-scale, long-range-aware features from each projection, enabling effective artifact suppression and robust 3D feature querying. To further enhance 3D spatial consistency, Trans$^2$-CBCT extends this design by integrating a Point Transformer equipped with learnable 3D positional encodings and a Neighbor-Aware Attention mechanism. This component enforces local smoothness and preserves anatomical details, particularly in highly undersampled settings. Extensive experiments on both LUNA16 and ToothFairy datasets demonstrate that our models consistently outperform analytic and learning-based baselines in terms of both PSNR and SSIM. Notably, Trans-CBCT achieves a strong balance between reconstruction quality and efficiency, while Trans$^2$-CBCT pushes the state of the art in fidelity. In summary, our work demonstrates the effectiveness of combining advanced 2D feature extraction with 3D geometric reasoning for CBCT reconstruction. We believe that Trans-CBCT and Trans$^2$-CBCT provide a strong foundation for future research on transformer-based volumetric imaging in sparse-view scenarios, with potential applications in low-dose medical imaging and mobile diagnostic systems.

{
    \bibliographystyle{IEEEtran}
    \bibliography{main}
}

\newpage

\vfill

\end{document}